\documentclass[12pt,a4paper,oneside]{article}

\usepackage{graphicx} 
\usepackage{psfrag} 
\usepackage{subfigure} 
\usepackage{array} 
\usepackage{booktabs} 
\usepackage{amsfonts} 
\usepackage{amssymb} 
\usepackage{amsmath} 
\usepackage{latexsym} 
\usepackage{amsthm} 
\usepackage{slashed} 
\usepackage{bm} 
\usepackage{mathrsfs} 
\usepackage[usenames,dvipsnames]{color} 
\usepackage{subfigure} 
\usepackage{paralist} 

\newcommand{\be}{\begin{equation}}
\newcommand{\ee}{\end{equation}}

\begin{document}

\title{Patterns of flavour violation \\ at the dawn of the LHC era}
\author{Maria Valentina Carlucci \\ \small{\emph{Physik-Department, Technische Universit\"at M\"unchen,}} \\ \small{\emph{James-Franck-Stra{\ss}e, D-85748 Garching, Germany}}}
\date{}
\maketitle

\begin{abstract}
The experimental success of the predictions of the Standard Model in the flavour sector suggests that New Physics should possess a highly non-generic flavour structure. Different approaches to this idea of Minimal Flavour Violation have been proposed and studied during the last years. At last, the LHC provides the possibility to test these patterns from different points of view, i.e.~through the direct search of New Physics at the high energy frontier, and through the indirect constraints at the intensity frontier of precision measurements.
\end{abstract}

\section{Introduction}
Despite the Standard Model (SM) provides an experimentally successful description of particle interactions, there are well known reasons, from the divergence of the Higgs mass to the unexplained phenomena as the dark matter or the baryon-antibaryon asymmetry, to believe that it represents only the low-energy limit of a more general New Physics (NP) theory. On one hand, theoretical arguments based on a natural solution of the hierarchy problem suggest that the cutoff energy scale $\Lambda$ should not exceed a few TeV. Another strategy to obtain clues about the value of $\Lambda$ is to build effective non-renormalizable interactions that encode the presence of new degrees of freedom at high energy: they are suppressed by inverse powers of $\Lambda$, and they can be constrained by experiments. For example, $\Delta F = 1$ and $\Delta F = 2$ flavour-changing neutral current (FCNC) effective operators could in general induce large effects in processes that are not mediated by tree-level SM amplitudes, and the absence of evidence of sizable deviations in the very abundant and precise data in $K$ and $B$ physics leads to bounds above $10^2-10^3$ TeV for the cutoff scale. The large discrepancy between these two determinations of $\Lambda$ is a manifestation of what in different specific NP frameworks (supersymmetry, technicolor, etc.) goes under the name of \emph{flavour problem} \cite{Isidori:2003qc}: it seems that NP presents the same flavour pattern as the SM, and this pattern has not been identified yet because the SM, which is the low-energy limit of the new theory, does not posses an exact flavour symmetry, but instead simply a flavour structure that could only be learnt from data. The formalization of this idea, i.e.~the recognition or imposition of the SM flavour structure in a certain NP model, is know as \emph{Minimal Flavour Violation} (MFV), and it has been developed and analyzed in different ways during the last years; here we present a review of the pragmatic approach called Constrained Minimal Flavour Violation (CMFV), and of a more formal approach that makes use of group methods and effective theories, and which at the same time allows more freedom.

The efforts performed during the last years in the theory and phenomenology of the flavour sector can be now finally supported from new data coming from the LHC, giving the possibility of obtaining more information about the flavour structure of the SM and beyond. Some of the most awaited measurements have represented another success for the SM: the CP-violating phase in the decay $B_s^0 \rightarrow J/\psi \; \phi$, previously suspected to show the presence of NP, has been found to be well consistent with the SM within current uncertainties \cite{LHCb:2011aa}, and  the new constraints on the decays $B^0_{(s)} \rightarrow \mu^+ \mu^-$, of great interest because of their theoretical clearness and their large NP allowance, are now very close to the SM prediction as well \cite{Aaij:2012ac}. Nevertheless, tensions of the order of 1-3$\sigma$ that have emerged in the last years in flavour observables are still present, like the in the determination of $\beta$ from $\epsilon_K-S_{\psi K_S}$ \cite{Lunghi:2010gv}, in the determination of $|V_{ub}|$ \cite{Beringer:1900zz}, in the branching ratio of $B \rightarrow \tau \; \nu$ \cite{Lunghi:2010gv}, in the like-sign dimuon charge asymmetry \cite{Abazov:2010hv}. All these data draw a non-trivial picture of the situation of the SM and beyond, and the time of putting it in a corner through precision measurements and calculations seems to be very close.

\section{Constrained Minimal Flavour Violation}

CMFV can be seen as a brute-force method of extrapolating the flavour structure of the SM. It is defined by two conditions \cite{Buras:2000dm}, \cite{Blanke:2006ig}:
\begin{itemize}
\item all flavour changing transitions are governed by the CKM matrix with the CKM phase being the only source of CP violation;
\item the only relevant operators in the effective Hamiltonian below the weak scale are those that are also relevant in the SM.
\end{itemize}

\subsection{Theory}

The standard tool for the study of weak decays of hadrons is the Operator Product Expansion \cite{Wilson:1969zs}, which permits to derive an effective low energy theory for the weak interactions of quarks, characterized by a scale $\mathcal{O}(M_W)$, when they are bound by strong interactions of the much lower hadronic energy scale of $\mathcal{O}(1 \text{ GeV})$. Using this technique the amplitude of a process can be calculated as
\be
A(I \rightarrow F) = \left\langle F \right| \mathcal{H}_{\text{eff}} \left| I \right\rangle = \frac{G_F}{\sqrt{2}} \sum_i V^i_{\text{CKM}} C_i (\mu) \left\langle Q_i (\mu) \right\rangle ~,
\ee
where $Q_i$ are the local operators for the Dirac structures that contribute to the process and the CKM factors $V^i_{\text{CKM}}$ and the Wilson coefficients $C_i (\mu)$ describe the strength with which they contribute. The scale $\mu$ separates the physics effects into short distance perturbative contributions, contained in $C_i (\mu)$, and the long distance contributions, contained in $\left\langle Q_i (\mu) \right\rangle$ and generally non-perturbative. This scale can be chosen arbitrarily, and it is standard to choose it of $\mathcal{O}(m_b)$ and $\mathcal{O}(1-2 \text{ GeV})$ for $B$ and $K$ decays respectively. However, if the aim is to expose the short distance structure of flavour physics and in particular the NP contributions, it is much more useful to choose a scale $\mu_H \sim \mathcal{O}(M_W, m_t)$ as high as possible but still low enough so that below it the physics is fully described by the SM \cite{Buras:2003jf}. Hence the relevant Wilson coefficients are be obtained as
\be
C_i (\mu) = \sum_j U_{ij} (\mu, \mu_H) C_j (\mu_H) ~,
\ee
where $U_{ij} (\mu, \mu_H)$ are the elements of the renormalization group evolution matrix, and the coefficients $C_j (\mu_H)$ can be found in the process of matching the full and the effective theory, and they will be a linear combination of certain loop functions $F_k$:
\be
C_j (\mu_H) = g_j + \sum_k h_{jk} F_k (m_t, \rho_{\text{NP}});
\ee
these functions will derive from the calculations of penguin and box diagrams containing the top quark and possible new heavy particles, and hence will depend on the parameters $\rho_{\text{NP}}$ of the NP model; the other SM contributions are contained in the constant term. As a consequence, the process amplitude will take the form
\be
A(I \rightarrow F) = P_0 (I \rightarrow F) + \sum_k P_k (I \rightarrow F) F_k (m_t, \rho_{\text{NP}}) ~,
\ee
where the coefficients $P_i$ collect different contributions:
\be
P_i (I \rightarrow F) \propto V^i_{\text{CKM}} B_i \eta^i_{\text{QCD}} ~,
\ee
where $V^i_{\text{CKM}}$ denote the relevant combinations of elements of the flavour matrix, $\eta^i_{\text{QCD}}$ stand symbolically for the renormalization group factors coming from $U_{ij} (\mu, \mu_H)$, $B_i$ are non-perturbative parameters representing hadronic matrix elements $\left\langle Q_i (\mu) \right\rangle$.

The advantages of this approach, known as \emph{penguin-box expansion} \cite{Buchalla:1990qz}, become evident when one considers the properties of the contributions involved.
\begin{itemize}
\item In the SM, since the only source of flavour and CP violation is the mass matrix, that has been factored out, the master functions $F_i$ are \emph{universal} (i.e., process independent), and \emph{real}.
\item As there are no right-handed charged current interactions, in the SM only certain number of local operators is present, and hence only a particular set of parameters $B_i$ is relevant. Similarly, if a careful treatment of the QCD corrections is performed, the factors $\eta^i_{\text{QCD}}$ can be calculated within the SM independently from the choice of the operator basis in the effective weak Hamiltonian. In conclusion, the $P_i$ coefficients are process dependent, but they depend only on the operator structure of the model.
\end{itemize}
As a consequence, the $P_i$ are \emph{model independent} within the whole class of CMFV, while the details of the single models are contained into the master functions $F_i$ that preserve their universality and realness. Since the SM  belongs itself to the class of CMFV, in these models the formulae of the observables will have the same form as in the SM with the only substitution $F_i (x_t) \rightarrow F_i (x_t, \rho_{\text{NP}})$. 

\subsection{Phenomenology}

\subsubsection{Universal Unitarity Triangle}

\begin{figure}
\begin{center}
\begin{minipage}[c]{.55\textwidth}
\centering
\includegraphics[width=\textwidth]{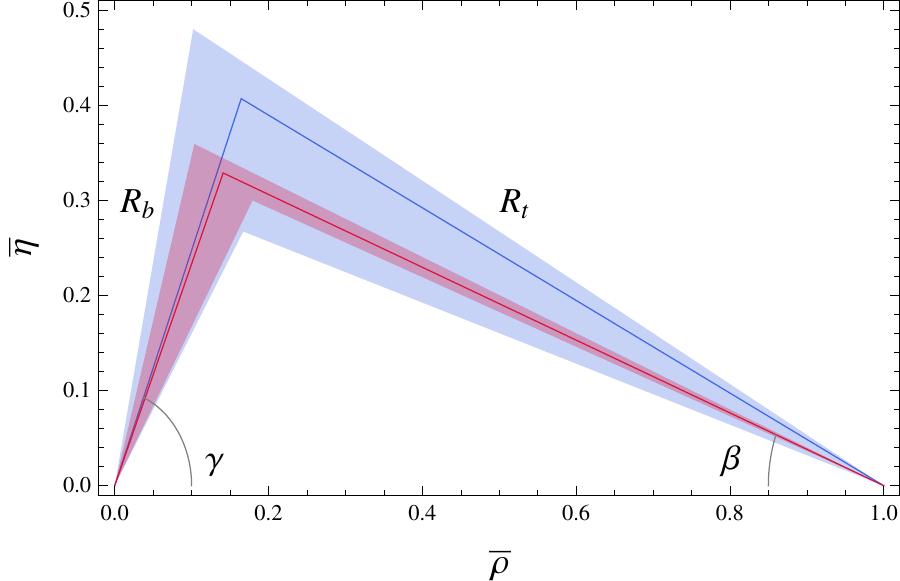}
\end{minipage}%
\hspace{.06\textwidth}%
\begin{minipage}[c]{.35\textwidth}
\centering
\begin{tabular}{ll}
\hline
\tiny{$|V_{us}|$} & \tiny{$0.2252 \pm 0.0009$} \\
\tiny{$|V_{cb}|$} & \tiny{$(4.9 \pm 1.1) \times 10^{-3}$} \\
\tiny{$|V_{ub}|_{\text{excl}}$} & \tiny{$(3.23 \pm 0.31) \times 10^{-3}$} \\
\tiny{$|V_{ub}|_{\text{incl}}$} & \tiny{$(4.41 \pm 0.15^{+0.15}_{-0.19}) \times 10^{-3}$} \\
\tiny{$\gamma$} & \tiny{$(68^{+10}_{-11})^{\circ}$} \\
\hline
\tiny{$\xi$} & \tiny{$1.237 \pm 0.032$} \\
\tiny{$m_{B_d}$} & \tiny{$5279.58 \pm 0.17$ MeV} \\
\tiny{$m_{B_s}$} & \tiny{$5366.77 \pm 0.24$ MeV} \\
\tiny{$\Delta M_d$} & \tiny{$(3.337 \pm 0.033) \times 10^{-10}$ MeV} \\
\tiny{$\Delta M_s$} & \tiny{$(116.4 \pm 0.5) \times 10^{-10}$ MeV} \\
\tiny{$S_{\psi K_S}$} & \tiny{$0.668 \pm 0.023$} \\
\hline
\end{tabular}
\end{minipage}
\end{center}
\caption{Comparison between the RUT (blue) and the UUT (red), obtained from the 2012 updated inputs  \cite{Beringer:1900zz}, \cite{Laiho:2009eu} in the table on the right. The shaded regions show the 1$\sigma$ uncertainties.}
\label{triangle}
\end{figure}

The analysis of the Unitarity Triangle (UT) provides a powerful test of the flavour pattern of CMFV in a model independent way; in fact, a triangle common to the whole class of CMFV models, known as \emph{Universal Unitarity Triangle} (UUT) \cite{Buras:2000dm}, can be built, and comparing it with the \emph{Reference Unitarity Triangle} (RUT), can give information not only on the possible presence of NP, but also about its flavour structure (Fig.~\ref{triangle}).

Once the parameters $\lambda \equiv |V_{us}|$ and $A = |V_{cb}|/\lambda^2$, unaffected by NP, have been determined, the determination of the apex $(\bar{\rho},\bar{\eta})$ requires the knowledge of one side and one angle of the triangle, provided the CKM matrix is unitary. Two choices of sets with different characteristics are possible.
\begin{itemize}
\item {\bf $\mathbf{R_b}$ and $\mathbf{\gamma}$ .} $R_b \propto |V_{ub}|$ is extracted by the inclusive and exclusive decays $B \rightarrow X_u \ell \bar{\nu}$, while $\gamma = \text{Arg} (V_{ub})$ comes from $B^{\pm} \rightarrow D \, K^{\pm}$; these are all tree-level processes, very unlikely modified by new physics. The triangle obtained with this method is the RUT.
\item {\bf $\mathbf{R_t}$ and $\mathbf{\beta}$ .} With a very good approximation $R_t \propto |V_{td}/V_{ts}|$, while $\beta = \text{Arg} (V_{td})$; the presence of the top quark implies that these can be only determined from loop processes. However, since in CMFV one simply has
\be
\Delta M_{d,s} = \frac{G_F^2}{6 \pi^2} \eta_B M_W^2 (\hat{B}_{B_{d,s}} F^2_{B_{d,s}}) m_{B_{d,s}} |V_{t(d,s)}|^2 F(x_t, \rho_{\text{NP}}) ~,
\ee
it is evident how the universality of the master function $F(x_t, \rho_{\text{NP}})$ implies a model independent extraction of the ratio
\be
\left| \frac{V_{td}}{V_{ts}} \right| = \xi \sqrt{\frac{m_{B_s}}{m_{B_d}}} \sqrt{\frac{\Delta M_d}{\Delta M_s}} ~,
\ee
where $\xi$ can be determined with non-perturbative methods and the other quantities are measurable. In an analogous way this ratio can also be extracted in for all the CMFV models from
\be
\left| \frac{V_{td}}{V_{ts}} \right|^2 = \frac{\mathcal{B}(B_d \rightarrow \mu^+ \mu^-)}{\mathcal{B}(B_s \rightarrow \mu^+ \mu^-)} ~.
\ee
On the other hand, the absence of new CP-violationg phases in the $B-\bar{B}$ mixing permits to write
\be
\sin 2 \beta = S_{\psi K_S}
\ee
and hence allows to determine the angle $\beta$ directly from the time-dependent asymmetry in the decay $B \rightarrow \psi K_S$. The triangle built in this way is common to all the models with CMFV and has been called UUT.
\end{itemize}

\subsubsection{Correlations and lower bounds}

The flavour universality of the master functions $F_i$ and the model independence of the parameters $P_i$ suggest the idea of considering ratios between different observables in which at the same time the flavour pattern can be tested and the hadronic uncertainties are reduced \cite{Buras:2003td}. Correlations like
\begin{subequations}
\be
\frac{\Delta M_d}{\Delta M_s} = \frac{M_{B_d}}{M_{B_s}} \frac{\hat{B}_d}{\hat{B}_s} \frac{F^2_{B_s}}{F^2_{B_s}} \left| \frac{V_{td}}{V_{ts}} \right|^2 r(\Delta M) ~,
\ee
\be
\frac{\mathcal{B} (B^0_s \rightarrow \mu^+ \mu^-)}{\mathcal{B} (B^0 \rightarrow \mu^+ \mu^-)} = \frac{\tau(B_s)}{\tau(B_d)} \frac{M_{B_s}}{M_{Bs}} \frac{F^2_{B_d}}{F^2_{B_s}} \left| \frac{V_{td}}{V_{ts}} \right|^2 r(\mu^+ \mu^-) ~,
\ee
\be
\frac{\mathcal{B} (B \rightarrow X_d \nu \bar{\nu})}{\mathcal{B} (B \rightarrow X_s \nu \bar{\nu})} = \left| \frac{V_{td}}{V_{ts}} \right|^2 r(\nu \bar{\nu})
\ee
\end{subequations}
have been indicated as \emph{standard candles of flavour physics} \cite{Buras:2012ts}, since in CMFV $r(\Delta M)=r(\mu^+ \mu^-)=r(\nu \bar{\nu})=1$ and deviations from unity can be used to recognize and parametrize different patterns of flavour violation.

Less intuitive but still simple calculations permit also to find that the observables related to the meson mixing like $\epsilon_K$ and $\Delta M_{d,s}$ are not only correlated, but also can only be enhanced with respect to the SM \cite{Blanke:2006yh}.

\subsection{Comparison with experiments}

Since a large discrepancy is present between the inclusive and exclusive determination of the CKM matrix element $V_{ub}$ \cite{Beringer:1900zz}, choosing one of the two values according to the impact on other observables can seems a criterion more meaningful than considering an average of the two. In the SM the exclusive measurement of $V_{ub}$ implies the prediction for the asymmetry $S_{\psi K_S}$ to be in agreement with data but $\epsilon_K$ to be below data, while the inclusive brings $\epsilon_K$ in the right range at the price of an enhancement of $S_{\psi K_S}$.

\begin{figure}
\centering
\subfigure{\includegraphics[width=0.48\textwidth]{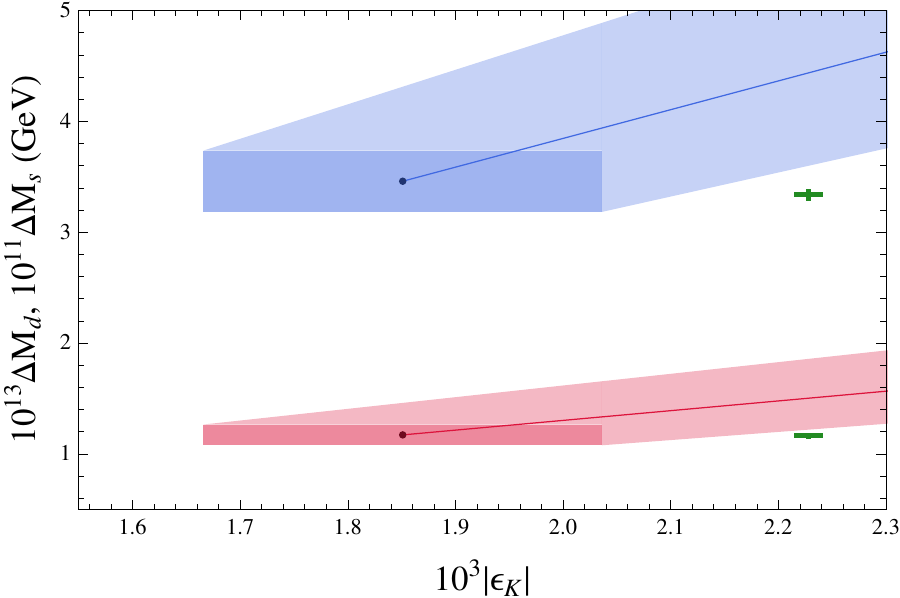}}
\quad
\subfigure{\includegraphics[width=0.48\textwidth]{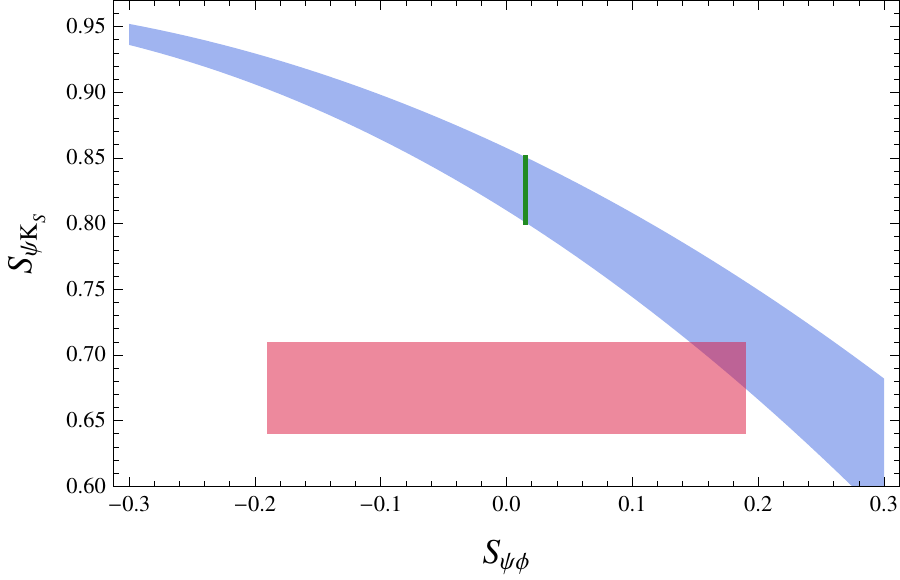}}\
\caption{(Left) $\Delta M_d$ (red) and $\Delta M_s$ (red) as functions of $\epsilon_K$ in models with CMFV. The green crosses represent the data, while the darker regions the SM predictions. (Right) Correlation between $S_{\psi K_S}$ and $S_{\psi \phi}$ in the 2HDM$_{\overline{\text{MFV}}}$, with a 2$\sigma$ uncertainty in blue. SM is represented by the green line, while the red area is the 2$\sigma$ experimental range.}
\label{cmfv2hdm}
\end{figure}

Since $S_{\psi K_S}$ cannot receive new contributions in models with CMFV, in this framework the exclusive value of $V_{ub}$ is preferred; on the other hand, as we have discussed, positive contributions to $\epsilon_K$ are allowed, and hence the $S_{\psi K_S}-\epsilon_K$ tension can be solved. Nevertheless, it has been shown that the enhancement of $\epsilon_K$ would determine a correlated enhancement of both $\Delta M_d$ and $\Delta M_s$, which are already slightly above the experimental values \cite{Buras:2012ts} (Fig.~\ref{cmfv2hdm}, left).

The previous considerations, even if only qualitative, point out the difficulties that CMFV models have in accommodating the tensions in flavour data, due to the presence of few free parameters and strict correlations. More quantitative studies, as well as a complete analysis of more observables, could be already able to derive more conclusive statements about the viability of this flavour violation scheme.

\section{Minimal Flavour Violation at Large}

\subsection{Effective field theory approach}

The largest group of unitary quark field transformations that commutes with the SM gauge Lagrangian is
\be
\mathcal{G}_q = \left(SU(3) \otimes U(1)\right)^3~,
\ee
i.~e.~a $SU(3)$ symmetry and a phase symmetry for each electroweak multiplet:
\be
SU(3)^3= SU(3)_{Q_L}\otimes SU(3)_{U_R} \otimes SU(3)_{D_R}~,  \; \; U(1)^3 = U(1)_B \otimes  U(1)_Y \otimes U(1)_{\rm PQ}~,
\ee
where the three $U(1)$ symmetries can be rearranged as the baryon number, the hypercharge, and the Peccei-Quinn symmetry. In the SM the Yukawa interactions break the flavour symmetry $\mathcal{G}_q$, but this can be formally recovered by promoting the Yukawa matrices to \emph{spurions}, i.~e.~dimensionless auxiliary fields transforming as 
\be
Y_u \sim (3, \bar 3,1)_{SU(3)^3}~, \qquad Y_d \sim (3, 1, \bar 3)_{SU(3)^3}~.
\ee
One defines an effective theory as satisfying the criterion of MFV if all higher-dimensional operators, constructed from SM and spurion fields, are formally invariant under the flavour group $\mathcal{G}_q$ \cite{D'Ambrosio:2002ex}.

In practice, one can build effective couplings and higher-dimensional operators in which the only relevant non-diagonal structures are polynomials $\mathcal{P}(Y_u Y_u^\dagger, Y_d Y_d^\dagger)$ of the two basic spurions
\be
Y_u Y_u^\dagger,~Y_d Y_d^\dagger \sim 
(8, 1,1)_{{\rm SU}(3)^3_q}\oplus(1, 1,1)_{{\rm SU}(3)^3_q}~.
\ee

As an example of this mechanism at work, we shortly discuss the application of this formulation of MFV to a generic model with two Higgs doublets, resulting in a NP scenario called  2HDM$_{\overline{\text{MFV}}}$ \cite{Buras:2010mh}.

\subsection{The 2HDM$_{\overline{\text{MFV}}}$}

\subsubsection{Imposing MFV}

In a generic  model with two-Higgs doublets, $H_1$ and $H_2$, with hypercharges $Y=1/2$ and $Y=-1/2$ respectively, the most general renormalizable and gauge-invariant interaction of them with the SM quarks is 
\be
- \mathcal{L}_Y = \bar Q_L X_{d1} D_R H_1 + \bar Q_L X_{u1} U_R H_1^c 
+ \bar Q_L X_{d2} D_R H_2^c + \bar Q_L X_{u2} U_R H_2 +{\rm h.c.}~,
\ee
where $H_{1(2)}^c = -i\tau_2 H_{1(2)}^*$  and the $X_i$ are $3\times 3$ matrices with a generic flavour structure. By performing a global rotation of angle $\beta = \text{arctan} (v_2/v_1)$ of the Higgs fields, the mass terms and the interaction terms are separated, but they cannot be diagonalized simultaneously for generic $X_i$ and dangerous FCNC couplings to the neutral Higgses appear.

If the MFV hypothesis is imposed instead, the $X_i$ are forced to assume a the particular structure
\begin{subequations}
\begin{align}
X_{d1} &= Y_d \\
X_{d2} &= \mathcal{P}_{d2}(Y_u Y_u^\dagger, Y_d Y_d^\dagger) \times Y_d = \epsilon_{0} Y_d + \epsilon_{1} Y_d  Y_d^\dagger Y_d                    
+  \epsilon_{2}  Y_u Y_u^\dagger Y_d + \ldots \\
X_{u1} &= \mathcal{P}_{u1}(Y_u Y_u^\dagger, Y_d Y_d^\dagger) \times Y_u = \epsilon^\prime_{0} Y_u + \epsilon^\prime_{1}  Y_u Y_u^\dagger Y_u 
+  \epsilon^\prime_{2}  Y_d Y_d^\dagger Y_u + \ldots \\
X_{u2} &= Y_u
\end{align}
\end{subequations}
that is renormalization group invariant \cite{Buras:2010mh}. At higher orders in $Y_i Y_i^\dagger$ FCNCs are generated, and in order to investigate them one can perform an expansion in powers of suppressed off-diagonal CKM elements, so that the effective down-type FCNC interaction can be written as
\be
\mathcal{L}_{\rm MFV}^{\rm FCNC} \propto {\bar d}^i_L \left[
\left( a_0 V^\dagger \lambda_u^2 V + a_1 V^\dagger \lambda_u^2 V  \Delta + a_2  \Delta V^\dagger \lambda_u^2 V \right) \lambda_d \right]_{ij}
 d^j_R~\frac{S_2 + i S_3}{\sqrt{2}} \; + \text{ h.c. }~,
\ee
where $\lambda_{u,d} \propto 1/v \; \text{diag} \left( m_{u,d},m_{c,s},m_{t,b} \right)$, $\Delta = \text{diag} \left( 0,0,1 \right)$, and the $a_i$ are parameters naturally of $\mathcal{O} (1)$; this structure shows a large suppression due to the presence of two off-diagonal CKM elements and the down-type Yukawas \cite{Buras:2010mh}, demonstrating explicitly how MFV is effective and natural. 

It is remarkable that the mechanisms of flavour and CP violation do not necessary need to be related: in MFV the Yukawa matrices are the only sources of flavour breaking, but other sources of CP violation could be present, provided that they are flavour-blind: this happens when the FCNC parameters $a_i$ are allowed to be complex, as well as for the phases that can be present in the Higgs potential.

\subsubsection{Comparison with experiments}

With a more detailed analysis the following relevant properties have been found \cite{Buras:2010mh}:
\begin{itemize}
\item the impact in $K$, $B$ and $B_s$ mixing amplitudes scales with $m_s m_d$, $m_b m_d$ and $m_b m_s$ respectively;
\item new flavour-blind phases can contribute to the $B$ and $B_s$ systems in the following way:
\be
S_{\psi K_S} = \sin (2 \beta + 2 \phi_{B_d}) ~, \quad S_{\psi \phi} = \sin (2 |\beta_s| - 2 \phi_{B_s}) ~,
\ee
with $\phi_{B_s} = (m_s/m_d) \phi_{B_d}$; they are not present in the $K$ system instead.
\end{itemize}

The previous observations imply that $\epsilon_K$ can receive only tiny new contributions while $S_{\psi K_S}$ could be in principle sizably modified; as a consequence, for the reasons that have been discussed in the previous section, the 2HDM$_{\overline{\text{MFV}}}$ selects the inclusive $|V_{ub}|$. However, in this framework a suppression of $S_{\psi K_S}$ would determine a correlated enhancement of $S_{\psi \phi}$, an effect that was considered very welcome until last year, when LHCb put an end to the hopes of new physics in the $B_s$ mixing phase, and that therefore puts this model in difficulty \cite{Buras:2012ts}. 

The flavour-blind phases of the Higgs potential imply instead $\phi_{B_s} = \phi_{B_d}$ \cite{Buras:2010zm}, and could be used to remove the $S_{\psi K_S} - \epsilon_K$ anomaly, but the size of $\phi_{B_d}$ that is necessary would imply in turn $S_{\psi \phi} > 0.15$, which is 2$\sigma$ away from the LHCb central value \cite{Buras:2012ts} (Fig.~\ref{cmfv2hdm}, right).

\section*{Acknowledgments}

I would like to warmly thank Andrzej Buras for giving me the possibility to attend this school, Barbara Guerzoni for her availability and kindness, and Roberto Preghenella for the pleasant company during the days and the nights in Erice and Favignana. This work has been supported in part by the Graduiertenkolleg GRK 1054 of DFG.


\end{document}